\documentclass[final,3xp,times,twocolumn]{elsarticle}

\usepackage[exponent-product=\cdot]{siunitx}
\usepackage{amsmath}
\usepackage{aas_macros}
\usepackage{amssymb}

\journal{Astroparticle Physics}

\begin{document}

\newcommand{\be}{\begin{equation}}
\newcommand{\ee}{\end{equation}}
\newcommand{\diff}{\ \mathrm{d}}
\newcommand{\sdiff}{\mathrm{d}}
\newcommand{\del}{\partial}
\newcommand{\unit}{\ \mathrm}
\newcommand{\unitb}{\mathrm}
\renewcommand{\vec}{\mathbf}
\newcommand*{\thead}[1]{\multicolumn{1}{c}{#1}}

\begin{frontmatter}



\title{Multi-band implications of external-IC flares}


\author[label1]{Stephan Richter\corref{cor1}\fnref{fn1}}
\ead{srichter@astro.uni-wuerzburg.de}
\cortext[cor1]{Corresponding author}
\fntext[fn1]{\textit{Tel.:}+49 931 3186956}

\author[label2]{Felix Spanier}

\address[label1]{Lehrstuhl f\"ur Astronomie, Universit\"at W\"urzburg, Emil-Fischer-Stra\ss e 31, D-97074 W\"urzburg, Germany}
\address[label2]{Centre for Space Research, North-West University, 2520 Potchefstroom, South Africa}

\begin{abstract}
Very fast variability on scales of minutes is regularly observed in Blazars. The assumption that these flares are emerging from the dominant emission zone of the very high energy (VHE) radiation within the jet challenges current acceleration and radiation models. In this work we use a spatially resolved and time dependent synchrotron-self-Compton (SSC) model that includes the full time dependence of Fermi-I acceleration. We use the (apparent) orphan $\gamma$-ray flare of \textit{Mrk501} during MJD 54952 and test various flare scenarios against the observed data. We find that a rapidly variable external radiation field can reproduce the high energy lightcurve best. However, the effect of the strong inverse Compton (IC) cooling on other bands and the X-ray observations are constraining the parameters to rather extreme ranges. Then again other scenarios would require parameters even more extreme or stronger physical constraints on the rise and decay of the source of the variability which might be in contradiction with constraints derived from the size of the black hole's ergosphere.
\end{abstract}

\begin{keyword}
Galaxies: active \sep relativistic processes \sep radiation mechanisms: non-thermal \sep Galaxies: jets \sep BL Lacertae objects: individual: Mrk501


\end{keyword}

\end{frontmatter}


\section{Introduction}\label{sec:intro}
Rapid flares on the scales of minutes have now been observed for several Blazars. The observed timescales often contradict the parameters obtained from fits of synchrotron-self-Compton (SSC)~\cite{1991ApJ...377..403C} models (e.g~\cite{2011ApJ...727..129A} in combination with~\cite{2011ICRC....8..171P}). For the case of \textit{PKS2155}~\cite{2007ApJ...664L..71A} the timescale in the frame of the black hole is in contradiction to the light-crossing time of the Schwarzschild-radius $R_S/c$ for the predicted mass. Moreover some of these flares appear to be without counterpart in the first bump of the SED, i.e. the leptonic synchrotron emission. It is, however, unclear whether this is always an intrinsic effect or due to observational limitations. 

Usually the discrepancy between lightcurves and steady-state parameters is circumvented via the argument of two separate emission zones. It is clearly possible to construct a parameter set yielding the required variability timescales, but it is impossible to constrain this scenario and its parameters by observations\footnote{With only one bump measured for each component's SED either the maximum electron energy or the cooling strength becomes ambiguous.}. Furthermore the steady state emission is often well described by a single component, requiring the assumption of a second component with comparable flux just for the time of a flare. In the case of a $\gamma$-ray orphan flare the additional difficulty of obscuring the X-ray flare arises. Setting the ratio between the inverse Compton and the synchrotron flux sufficiently high will lead to strong IC-cooling and hence quite extreme parameters. Models that disconnect the source of the variability from the black hole were presented by e.g.~\cite{2012ApJ...749..119B}. Especially the case of the inverse Compton scattering of external radiation can produce an orphan flare, if the boost between the external source and the frame of the high energy particle distribution is sufficiently high. A possible origin of this external radiation and its effect on the steady state can be found in~\cite{2005A&A...432..401G}.

In any case a constriction of models explaining rapid variability self-consistently, i.e. emerging from the same spatial region as the steady state emission, is only possible via time dependent numerical modeling that is not constrained by light crossing times. The non-linear effects that will arise in such models were studied in~\cite{2012AIPC.1505..660Z}. However, no time-dependent solution of the system was given. Furthermore this work can not describe variability shorter than the light crossing time of the emission region due to the assumption of homogeneity.

A spatially and time-dependent model is presented in section~\ref{sec:model}. In section ~\ref{sec:results} we then present a fit of the averaged spectral energy distribution (SED) of \textit{Mrk501} in 2009 obtained from~\cite{2011ApJ...727..129A} and test various flare scenarios and their multi-band implications against the fast flare with $t_{var}\approx\SI{2000}{s}$~\cite{2011ICRC....8..171P,2011ICRC....8...78B}. The physical constraints of these scenarios as well as an ambiguity of the SED fitting and its consequences for the time evolution is discussed in section~\ref{sec:discussion}.

\section{Model}
\label{sec:model}
\begin{figure}[ht]
  \centering
  \includegraphics[width=0.5\textwidth]{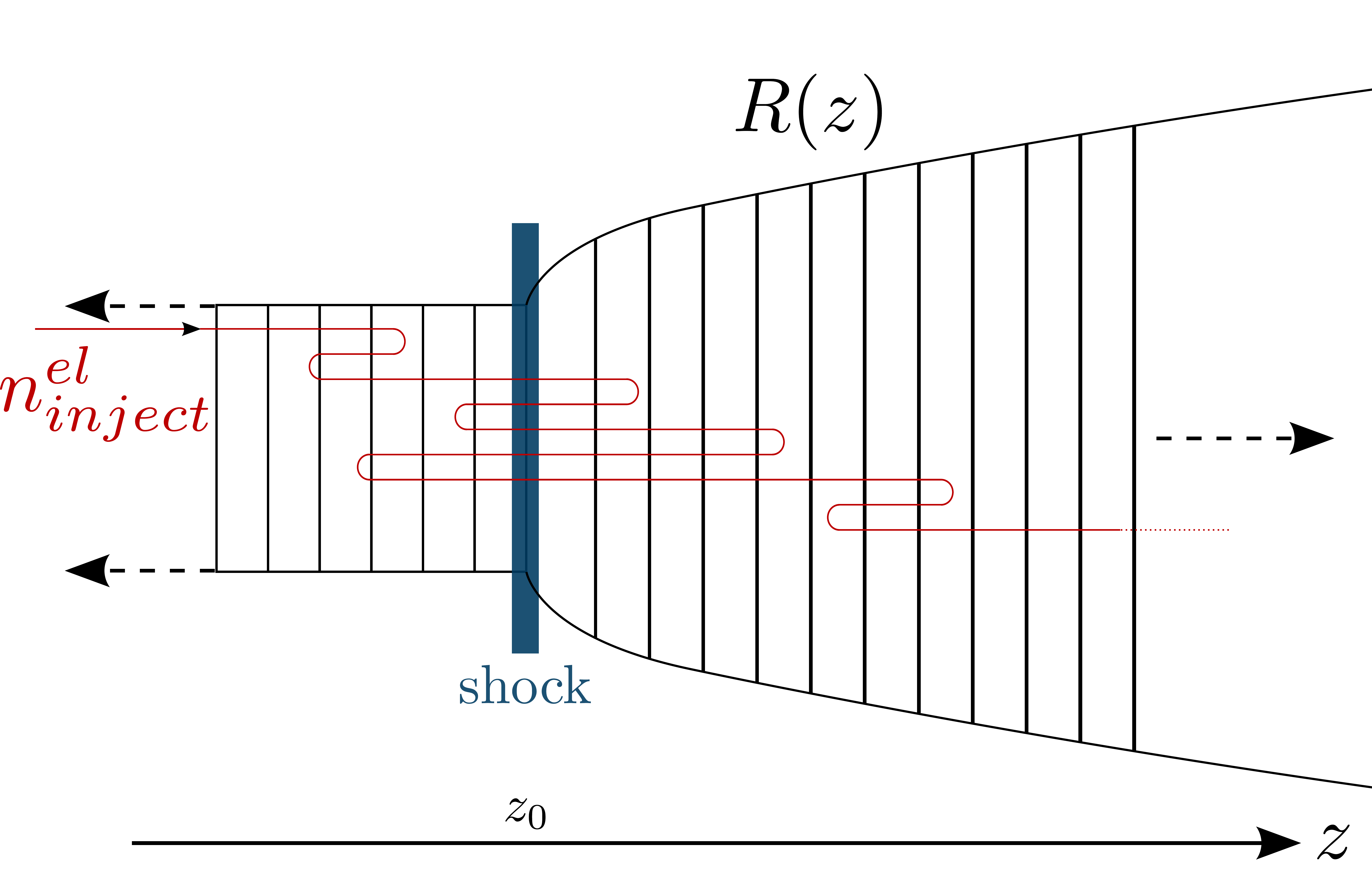}
  \caption{A schematic illustration of the used geometry. The spatial discretization is performed along $z$, the shock normal and parallel to the magnetic background field. A shock is represented as a jump in velocity of the background plasma. Particles can move in both directions and scatter between them. The scattering is elastic in the rest frame of the background plasma.}
  \label{fig:geometrie}
\end{figure}

The numerical model used in this work connects the dominant acceleration process, which is Fermi-I shock acceleration~\cite{1983ApJ...270..319W}, with the geometry of the simulation box. For this purpose we use a spatial discretization along the shock-normal. A shock can then be represented by a jump of the bulk velocity of the ambient plasma between neighboring cells. Particles can convect back and forth and are pitch-angle-scattered at a rate that is parameterized by the isotropisation timescale $t_{iso}$. Hence the standard picture of the Fermi-I process (e.g. \cite{2004PASA...21....1P}) evolves naturally. This is roughly sketched in Fig.~\ref{fig:geometrie}

For this work we used the simplest setup with $R(z)=R_0$. This implies no cooling due to adiabatic expansion and a constant magnetic field $B(z)=B_0$. The only inhomogeneity arises from the presence of the shock, which will produce a jump in the velocity of the bulk plasma. A more elaborated setup could be achieved by a spatially dependent scattering rate (e.g. approaching the shock) or a strong expansion behind the shock to introduce a natural boundary condition. Such setups are, however, not relevant for the aim of this study and would introduce additional difficulties. The physics of the pre-shock region are hardly understood even for weak solar shocks and would introduce additional parameters in the here presented case. The effect of expansion behind the shock on the boundary conditions can be found in~\cite{2013EPJWC..6105010R}. For the studies of the variability of the upper energy range of both bumps in the SED (i.e X-rays and TeV-range) an artificial boundary due to a limited size of the emission region is sufficient. The only condition, that has to be fulfilled, is the escape time $t_{esc}$ to be larger than the variability time scale $t_{var}$. Otherwise timescales would be lowered artificially. In our geometry this is true if $t_{var}<z_{max}/V_P=z_{max}\cdot r/V_S$\footnote{Note that $z_{max}$ is specified in the shock-frame.}.

The shock velocity $V_S$ and its compression ratio $r$ is used to calculate the downstream velocity behind the shock $V_P$ (in units of $c$), expressed in the upstream frame:
\be
  V_P=\frac{V_S(r-1)}{r-V_S^2}
\ee

In order to connect this intrinsic inhomogeneity with the acceleration of particles, pitch angle scattering and the resulting acceleration (in the shock frame) has to be modeled. Therefore a distinction between the particles that move towards the downstream and those that move towards the far upstream is introduced. Details of this modeling can be found in~\cite{2012IJMPS..08..392R}. The computation of the radiative output is then performed in the isotropic approximation.

\subsection{Kinetic Equations}
In addition to the Fermi-I acceleration the particle distribution $n(\gamma,t)$ will be determined by Fermi-II acceleration (momentum diffusion coefficient $D=\frac{v_A^2}{9\kappa_\parallel}$, where $v_A$ is the Alfv\' en speed), the source term $S(z,\gamma,t)$ as well as losses due to synchrotron emission and inverse Compton scattering. The two latter processes are computed via
\be
	\label{eq:synclosses}
	P_{sync}=\frac{1}{6\pi}\frac{\sigma_TB^2}{mc}\gamma^2=\beta_s\gamma^2\quad,
\ee
and
\be
  \label{eq:pic}
  P_{IC,\gamma}=\frac{1}{mc^2}\int\sdiff\nu'\ h\nu'\int\sdiff\nu\ N(\nu)\frac{\sdiff N_{\gamma,\nu}}{\sdiff\nu'\diff t}\quad,
\ee
respectively. The time evolution of the electron density in the energy domain can then be computed via the Fokker-Planck equation integrated over $\mu$:
\begin{multline}
  \label{eq:gamma_kinetic_el}
  \frac{\del n(\gamma)}{\del t}=\frac{\del}{\del \gamma}\Biggl[D\gamma^2\cdot\frac{\del n(\gamma)}{\del\gamma} \\
  +(\beta_s\gamma^2-2D\gamma+P_{IC}(\gamma))\cdot n(\gamma)\Biggr]+S(z,\gamma,t)
\end{multline}

From Eq.~\ref{eq:pic} it is obvious that the inverse Compton process depends on the photon density and hence introduces a back reaction on the electron distribution. In order to solve the resulting non linear system in a time dependent way it becomes necessary to calculate the photon density simultaneously. The equation for $N(\nu,t)$ governing the time evolution is
\be
  \label{kinetic_ph}
  \frac{\del N(\nu)}{\del t}=-c\cdot\kappa_{\nu,SSA}\cdot N(\nu)+\frac{4\pi}{h\nu}\cdot(\epsilon_{\nu,IC}+\epsilon_{\nu,sync})-\frac{N(\nu)}{t_{esc}}\quad.
\ee
The dominant gain $\epsilon_{\nu,sync}$ due to synchrotron emission is computed in the Melrose-approximation~\cite{1983Ap&SS..92..105B}:
\be
  \label{eq:melrose}
  P_\nu(\gamma,\nu)\approx1{,}8\frac{\sqrt3\ q^3B}{m\ c^2}\cdot\left(\frac{\nu}{\nu_c}\right)^\frac{1}{3}\cdot e^{-\frac{\nu}{\nu_c}}
\ee
The synchrotron self absorption coefficient $\kappa_{\nu,SSA}$ is computed similarly.
The contributions of the IC process are calculated via
\be
  \label{eq:icwins}
  \epsilon_\nu=\frac{h\nu}{4\pi}\int\sdiff\gamma\ n(\gamma)\int\sdiff\nu'\left(\frac{\sdiff N_{\gamma,\nu'}}{\sdiff\nu\diff t}\cdot N(\nu')-\frac{\sdiff N_{\gamma,\nu}}{\sdiff\nu'\diff t}\cdot N(\nu)\right)\quad.
\ee
The escape timescale $t_{esc}$ includes the catastrophic losses. The computation of photon absorption due to pair production is implemented in our code, but is not relevant in this case. This can be seen using the delta-approximation for the cross-section (e.g.~\cite{1997A&A...325..866B}) and the values $\nu_c$ and $E_{ph,inj,0}$ in section~\ref{sec:externalcompton}. These result in an absorption coefficient $\alpha_{\nu\nu}\sim\SI{E-20}{cm^{-1}}\ll z_{max}^{-1}$.

The time dependent, over all flux from the complete simulation box is computed - following Blandford and K\"onigl~\cite{Blandford1979} - as an integral over all positions $z$, hence different light travel times to the observer are incorporated. The number of free parameters in the model is seven (magnetic field $B$, injection energy $\gamma_{inj}$, injection rate $\dot N_{el}$, size of the blob $R$, Doppler factor $\delta$, the acceleration efficiency and the shock compression ratio $r$). Furthermore the shock is described by the shock velocity. However, this parameter is well constrained by theory and has only a small influence on the resulting SED.

\section{Results}
\label{sec:results}
We use the model presented in section~\ref{sec:model} to obtain the steady state parameters for the emission region. This state is used as the starting point for the flare scenarios. The physical plausibility of these scenarios is discussed in section~\ref{sec:discussion}.

\subsection{Steady state}
\label{sec:steadystate}
\begin{figure}[ht]
 \centering
 \includegraphics[width=0.5\textwidth]{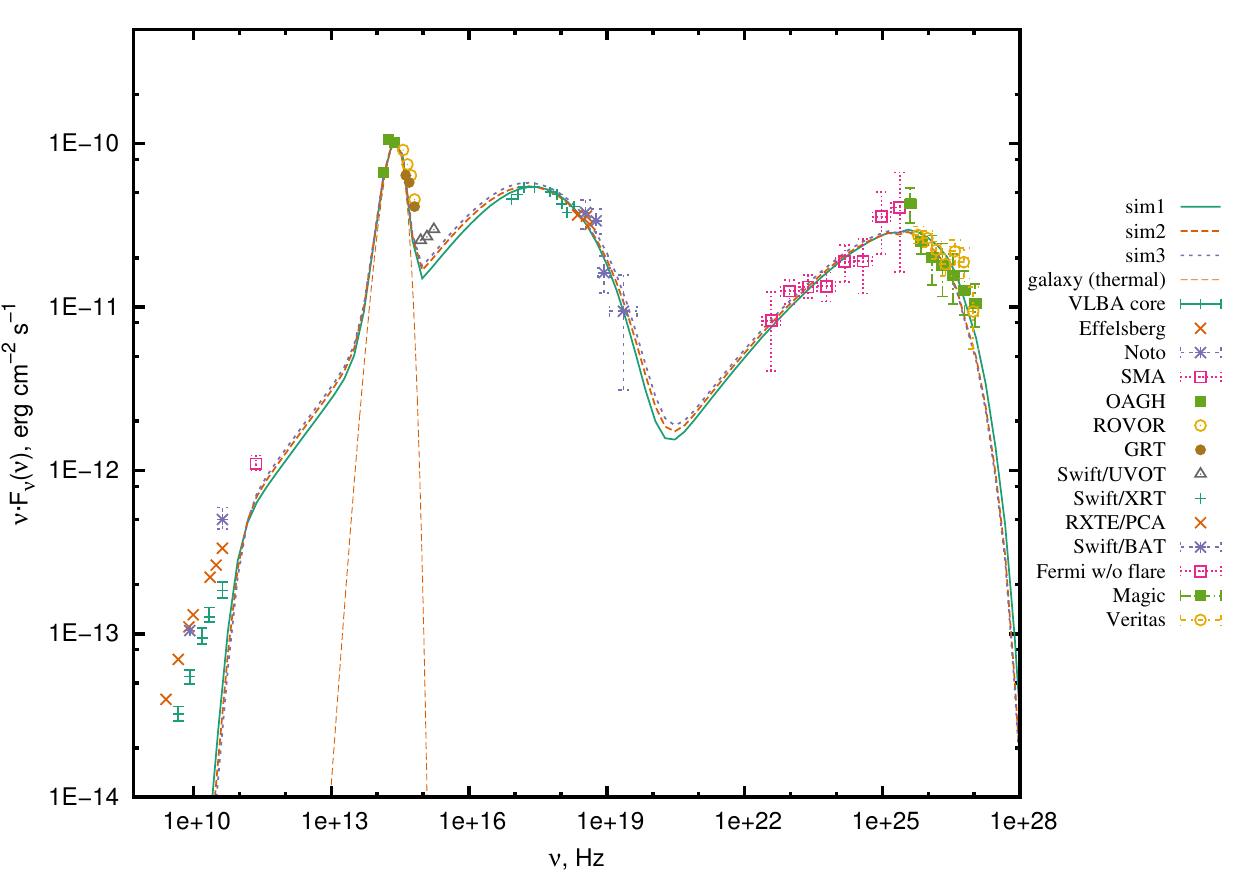}
\caption{Fit of the data obtained during a multi-frequency campaign~\cite{2011ApJ...727..129A}.}
\label{fig:steady_state_sed_fit}
\end{figure}

The fits of the averaged low state of this source were produced with the model described in section~\ref{sec:model} and are shown in Fig.~\ref{fig:steady_state_sed_fit}.
A good overall fit can be achieved, although the spectral index indicated by the \textit{UVOT} and \textit{SMA} data does not match the \textit{Fermi} spectral index. This problem was already addressed by~\cite{2011ApJ...740...64L,2011ApJ...743L..19L}. The flux measured by \textit{UVOT} could also emerge from the broad line region (BLR)~\cite{2012MNRAS.420.2899G}. The parameters summarized in table~\ref{tab:parameters} for \textit{sim1} differ from those presented by \cite{2011ApJ...727..129A}.
\begin{table}[h]
	\centering
	\caption{Parameters obtained from the fits in Fig.~\ref{fig:steady_state_sed_fit}.}
	\tabcolsep=0.14cm
	\begin{tabular}{cccccc}
		\hline\hline
		\thead{sim} & \thead{$z\ (cm)$} & \thead{$B\ (G)$} & \thead{$N_{inj}\ (s^{-1}$)} & \thead{$\delta$} & \thead{$t_{iso}\ (s)$}\\
		\hline
		\\[-3mm]
		sim1 & $4.5\cdot10^{15}$ & $0.044$ & $8.9\cdot10^{43}$ & $33.5$ & 5000\\
		sim2 & $3.8\cdot10^{15}$ & $0.06$ & $8.25\cdot10^{43}$ & $32.1$ & 125\\
		sim3 & $3.3\cdot10^{15}$ & $0.07$ & $8.25\cdot10^{43}$ & $32.1$ & 75\\
	\end{tabular}
	\label{tab:parameters}
\end{table}
The very large size of the emission region there is not compatible to the position of the cooling break for the given magnetic field. Therefore self-consistent (i.e. without a priori assumptions on the shape of the electron distribution) models can not produce such a fit. However, the values obtained by our fit are much less constraining for the presence of fast variability. The significance of the parameters most relevant for variability, magnetic field $B$ and acceleration timescale $t_{acc}\propto t_{iso}$ can be pushed even further, as shown in \textit{sim2} and \textit{sim3}. The limitation of this course is discussed in section~\ref{sec:discussion}.
The fitting was done ``by eye'', hence no significance for the resulting parameters was computed.

The parameters result in a light crossing time of $t_{lc}=\SI{4545}{s}$. We want to stress that in the case ($t_{var}<t_{lc}$) numerical studies of the variability are only possible with a spatially resolved model which preserves causality.

\subsection{Flare scenarios}
\begin{figure}[ht]
 \centering
  \includegraphics[width=0.5\textwidth]{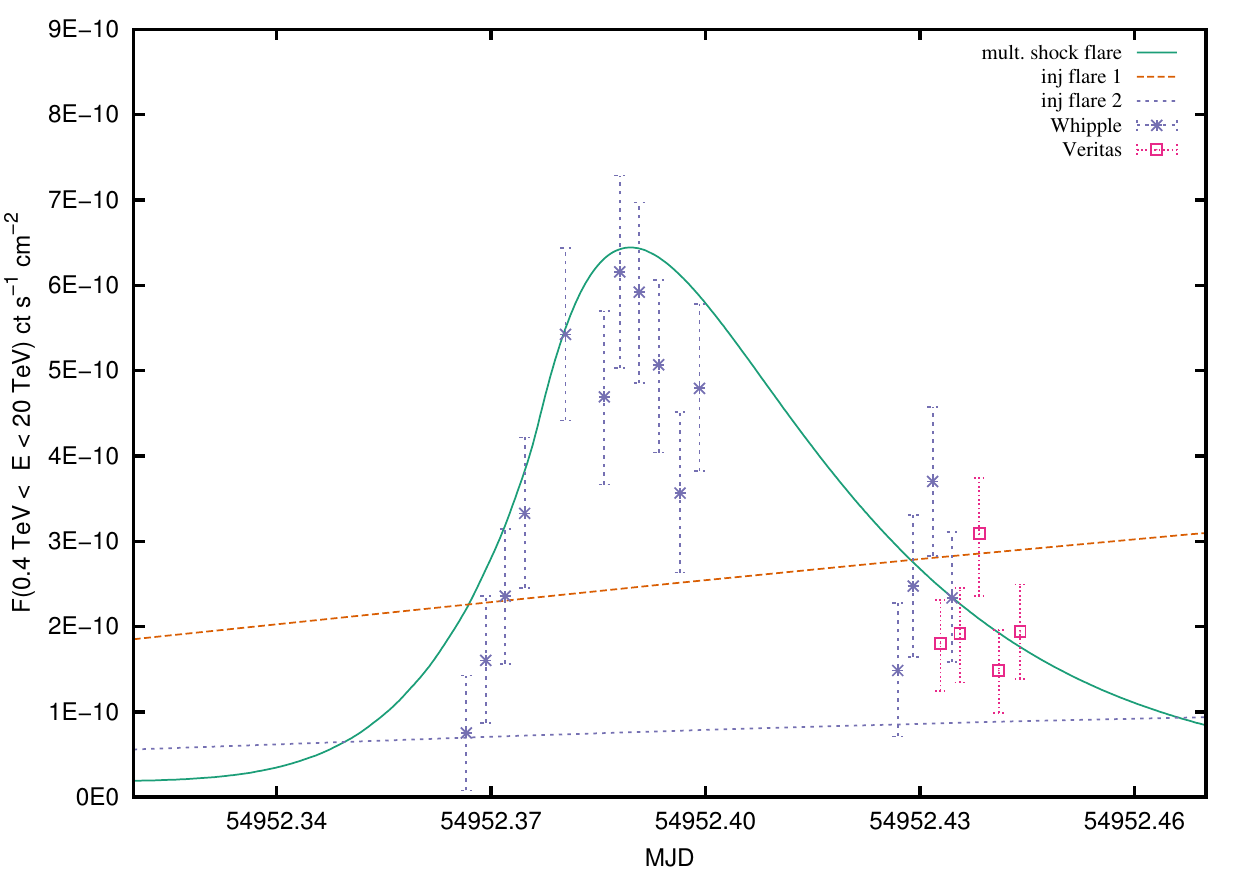}
\caption{Lightcurves produced for different flare scenarios compared with the observation by \textit{Whipple}~\cite{2011ICRC....8..171P}.}
\label{fig:lightcurve_veritas_rest}
\end{figure}

In the following we use simulation \textit{sim3} as a starting point for different flare scenarios. These are two different variations of the particle injection rate and a so called multiple shock scenario. In the latter an additional shock is present in the downstream that reaccelerates the power-law distribution. From Fig.~\ref{fig:lightcurve_veritas_rest} it can be seen that additional particle injection can not account for the observed variability timescale in the highest energies. The acceleration of particles from the injection energy up to the TeV scale on the observed timescale would require a much larger acceleration rate that is not compatible with the steady state SED, as mentioned in section~\ref{sec:steadystate}
An additional difficulty is the increase of inverse Compton cooling with higher particle densities. An increase by one order of magnitude, as in scenario \textit{inj flare 1}, will lead to a flux reduction in the \textit{Whipple}-band, unless the magnetic field is reduced simultaneously. This is on the one hand in contradiction with the equipartition assumption, on the other hand a lower magnetic field will further increase the cooling timescale.

\begin{figure}[ht]
 \centering
 \includegraphics[width=0.5\textwidth]{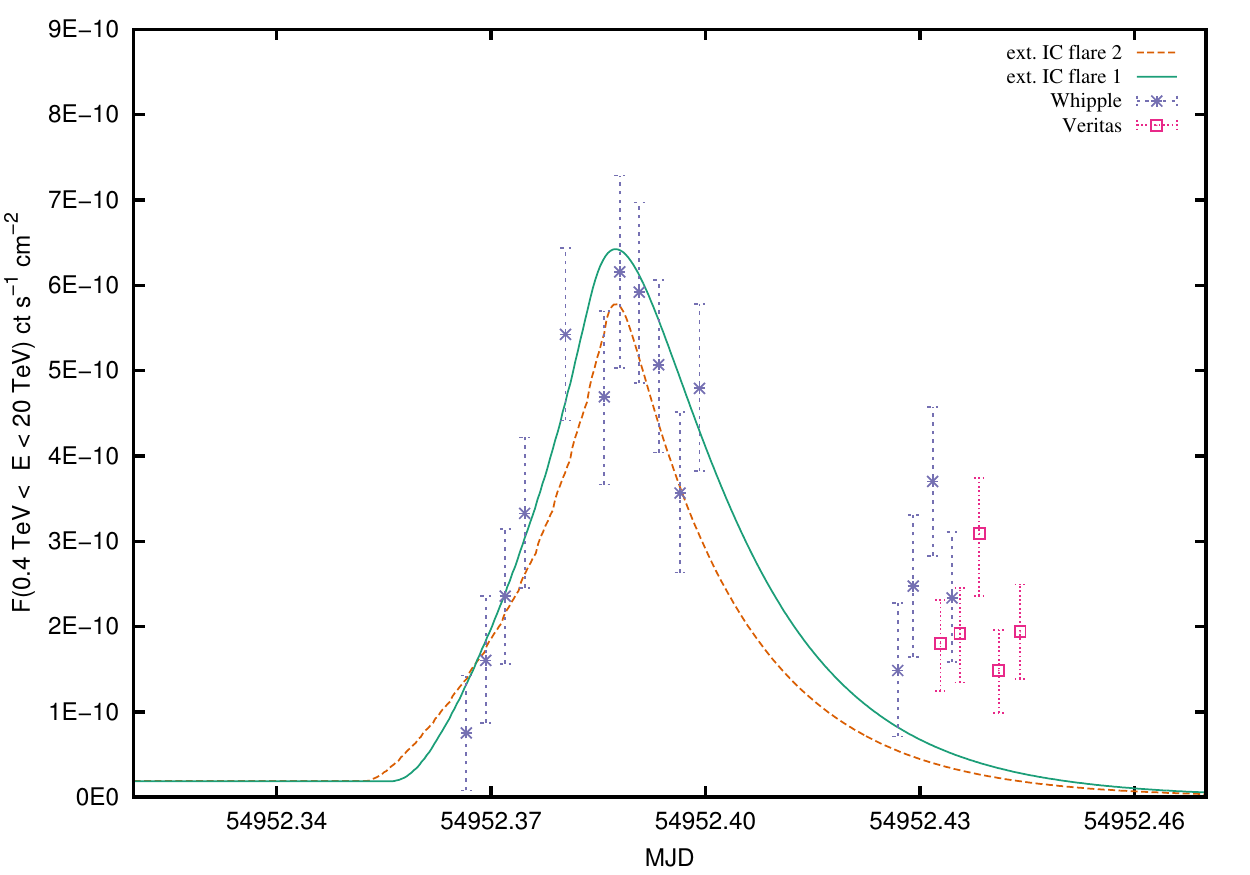}
\caption{Fits of the lightcurve recorded by \textit{Whipple}~\cite{2011ICRC....8..171P} employing the external inverse Compton scenario.}
\label{fig:lightcurve_fit}
\end{figure}

In the multi-shock-scenario only particles with energies above the cooling break are effected, which is, close to the primary shock, only a narrow band. Consequently the system returns to the steady state very fast, after the secondary shock disappears. This disappearance is however artificially introduced.
\paragraph{External inverse Compton scenario}\mbox{}\\
\label{sec:externalcompton}
Starting from the steady state fits, the lightcurves in Fig.~\ref{fig:lightcurve_fit} were obtained by injection of an external photon distribution.
A synchrotron spectrum was chosen for the spectral shape and approximated according to Eq.~\ref{eq:melrose}. The two occurring parameters are the cut off frequency and the normalization in the form of the differential energy density $E_{ph,inj,0}$ at $\nu_c$. For both \textit{sim2} ($\rightarrow$ \textit{IC flare 2}) and \textit{sim3} ($\rightarrow$ \textit{IC flare 1}) fits were produced.
The two parameters were set to $\nu_c=\SI{E16}{Hz}$, $E_{ph,inj,0}=\SI{6.6E-17}{erg s cm^{-3}}$ for \textit{IC flare 1} and $\nu_c=\SI{E15}{Hz}$, $E_{ph,inj,0}=\SI{3.3E-16}{erg s cm^{-3}}$ for \textit{IC flare 2}, respectively.

In contrast to the previously discussed scenarios, here the falling flank of the lightcurve is independent from the length of the injection and completely determined by inverse Compton cooling. In addition to the possible boosting this significantly relaxes the constraints on the source of the variability.

\subsection{Multi band view}
\label{sec:multiband}
\begin{figure}[ht]
 \centering
  \includegraphics[width=0.5\textwidth]{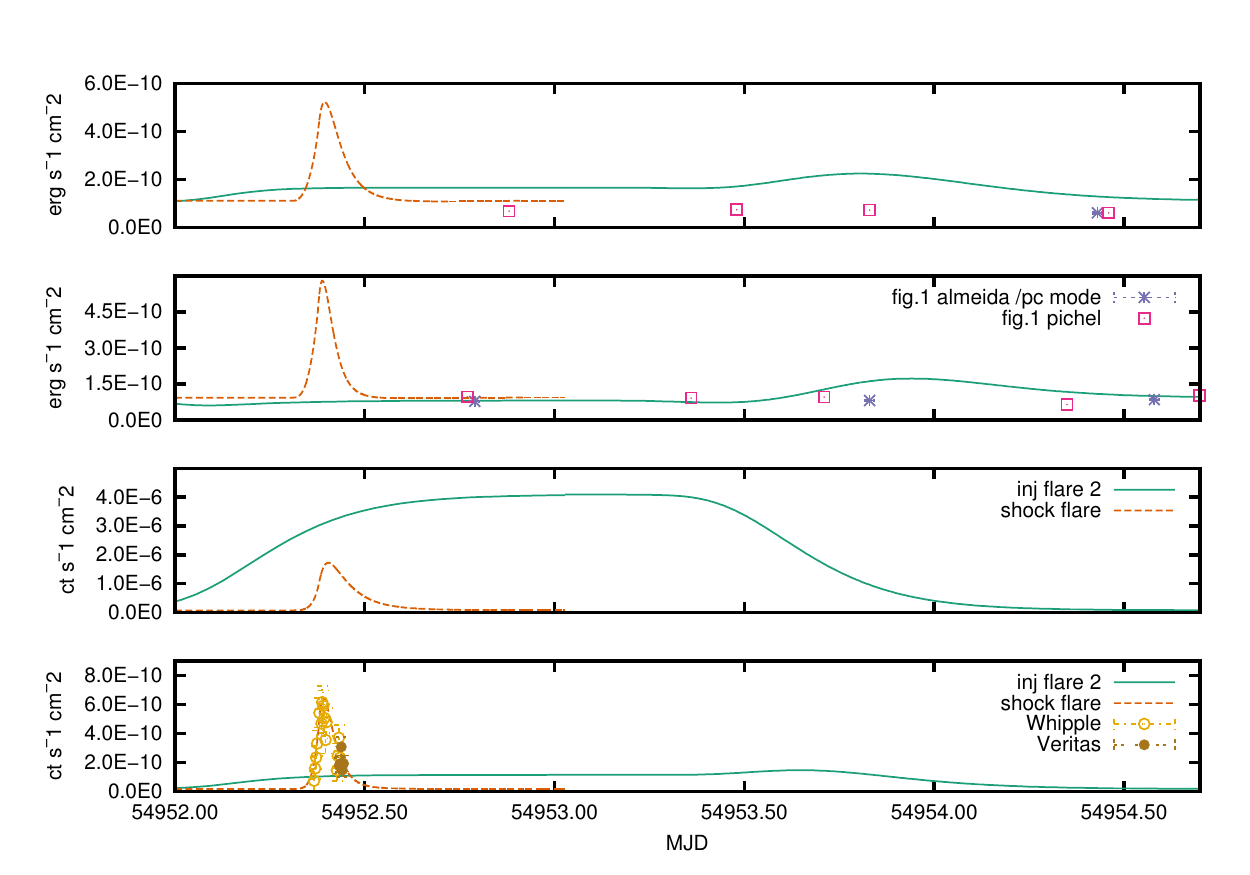}
\caption{Simulated lightcurves in various energy ranges for particle injection and the multi shock scenario.}
\label{fig:lightcurve_rest_multi}
\end{figure}

All presented flare scenarios produce variability in all relevant bands. An increase of the particle injection as well as the multiple shock scenario leads to a simultaneous increase in all bands, shown in Fig.~\ref{fig:lightcurve_rest_multi}. Only the hard X-ray flux stays approximately constant due to inverse Compton cooling in the injection case.
Here it is obvious, that even a magnetic field of almost $\SI{0.1}{G}$ can neither explain the rapid rise and fall in the TeV range, nor the absence of an increased X-ray flux, approximately $\SI{6}{h}$ after the TeV-flare. The falling flank of the shock-scenario is sufficiently steep.

\begin{figure}[ht]
 \centering
  \includegraphics[width=0.5\textwidth]{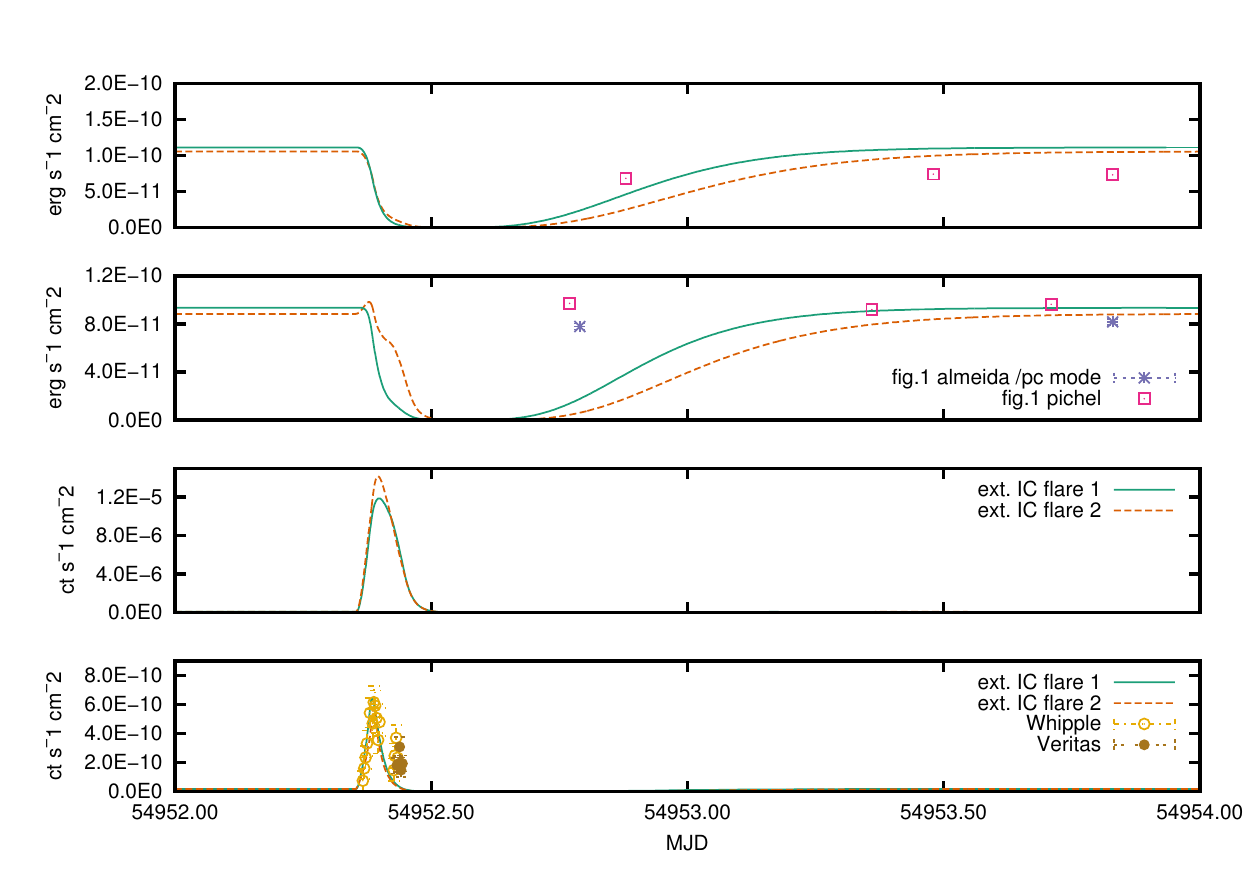}
\caption{Simulated lightcurves in various energy ranges for the external inverse Compton flares.}
\label{fig:lightcurve_ec_multi}
\end{figure}

The injection of a strong, additional photon field has an opposite effect on the electron synchrotron emission. As shown in Fig.~\ref{fig:lightcurve_ec_multi} the X-ray emission is reduced as long as the external field in present. For the presented simulations this timespan is $t_{inj}\approx\SI{1.5E4}{s}$ in the observers frame.
After stopping the injection the system returns to its steady state on the intrinsic timescale. It can be seen that \textit{sim3} is slightly faster, but can not explain the observed X-ray flux.
\section{Discussion}
\label{sec:discussion}
In this work we want to derive constraints on flare scenarios starting from a time averaged SED which was taken as the steady state. The parameters that dominantly determine the time evolution are the magnetic field $B$ (cooling timescale) and the isotropisation timescale $t_{iso}$ which is proportional to the acceleration time scale. These two parameters will also determine the cut-off frequencies of the two bumps of the SED. The maximum electron energy $\gamma_{max}$ is balanced between acceleration and synchrotron cooling:
\be
	\gamma_{max}\propto t_{acc}^{-1}B^{-2}
\ee
The cut-offs of the synchrotron spectrum and the IC-bump (in the Klein-Nishina regime) are then given by:
\begin{subequations}
\begin{align}
        \nu_{sync}^{max}&\propto\gamma_{max}^2B\propto t_{acc}^{-2}B^{-3}\\
        \nu_{IC}^{max}&\propto\gamma_{max}\propto t_{acc}^{-1}B^{-2}
\end{align}
\end{subequations}
These dependencies can be reproduced well with our code. Consequently it is not possible to set the values of these parameters to arbitrary high $B$ and low $t_{acc}$. However, due to the larger error bars in the falling flanks of the SED, the values can vary slightly. We find that values of $B>\SI{0.1}{G}$ can not be reached. As shown in section~\ref{sec:results} these parameter constraints prevent the reproduction of the very fast variability in the TeV band by a simple variation of the particle injection rate.

The injection of an external radiation field suffers a similar problem. This scenario leads, on the one hand to a rise in the very high energy range (VHE), on the other hand to a drop in the X-rays. The latter is due to the inverse Compton cooling of the electrons. Once this process dominates, the VHE flux reaches its maximum and starts to decrease. This relaxes the constraints on the source of the external radiation field. Assuming this source has a very small Doppler factor in the observers frame, the necessary timescales can be approximated as follows: The decay time in the frame of the external radiation source is $\widetilde{t}_{decay}=\frac{\delta^2}{\delta_{ext}}t_{inj}\approx\SI{1.5E7}{s}$. Here the Doppler factor of the emission blob $\delta$ was taken from table~\ref{tab:parameters} and that of the external source was set to $\delta_{ext}\approx1$. The rise time of the flare can be converted similarly to the variation timescale of the external source $\widetilde{t}_{var}\approx\SI{5E5}{s}$. For various scenarios like a stationary, flaring component within the jet~\cite{decelerating_flow}, a jet-star-interaction~\cite{2012ApJ...749..119B} or a different jet layer~\cite{spine_layer} it should be possible to accommodate these values.

The increase of the X-ray flux is however determined by the steady state parameters. To explain the unaltered X-ray flux after the TeV-flare, much more extreme parameters would be required that again are not in agreement with the steady state SED. Although this scenario can explain the narrow peak of the TeV-flux quite elegantly, it has to be neglected at least for this particular case.

The multiple shock scenario is the only one that can qualitatively explain the TeV lightcurve, while not being in contradiction with the X-ray data. An increased flux in this band can be obscured due to the very short flare time and the not strictly simultaneous observations. The time in which the second shock is present in the downstream of the primary shock is however introduced arbitrarily. Especially the time (and hence distance to the first shock) in which the shock has to decay is crucial for the shape of the lightcurve, that otherwise would be significantly higher and wider.
\section{Conclusion}
We presented numerical simulations for various self-consistent scenarios for the apparent orphan flare of \textit{Mrk501} during MJD 54952. It was shown that within this physical framework, i.e. the flare is emerging from the same region as the steady state emission is, a simple variation of particle density can not explain the observed lightcurves.

More elaborated scenarios, like the multiple shock and external inverse Compton cases, can explain the observed time scale and apparent orphan character of the flare. We also showed, that in a full, time dependent treatment in general every variation of physical parameters will lead to a change in flux in all relevant bands. Therefore simultaneous, in the case of extremely short flares even strictly simultaneous, observations are needed to distinguish possible scenarios.

In the case investigated in this work, the external inverse Compton flare can be ruled out, while the multiple shock scenario is not in contradiction with the data. A physical sound motivation for the appearance and disappearance of a second shock in the downstream of the acceleration region is however not on hand.\\
\textit{Acknowledgments} We thank the referee for his helpful comments and a detailed discussion. SR wants to thank GK 1147 for their support.





\bibliographystyle{model1-num-names}
\bibliography{apj-jour,ref}







\end{document}